\documentclass[11pt]{article}
\usepackage{geometry}                
\usepackage[parfill]{parskip}    
\usepackage{graphicx}
\usepackage{amssymb}
\usepackage{epstopdf}

\long\def\symbolfootnote[#1]#2{\begingroup%
\def\thefootnote{\fnsymbol{footnote}}\footnote[#1]{#2}\endgroup}

\title{Dark Matter Densities during the Formation of the First Stars and in 
Dark Stars}
\author{
\mbox{Katherine Freese$^1$\footnote{Email:ktfreese@umich.edu},}
\mbox{Paolo Gondolo$^2$\footnote{Email:paolo@physics.utah.edu},}
\mbox{J. A. Sellwood$^3$\footnote{Email:sellwood@physics.rutgers.edu},}
and
\mbox{Douglas Spolyar$^4$\footnote{Email:dspolyar@physics.ucsc.edu},}
}

\begin{document}
\maketitle

\begin{center}
$^1$Michigan Center for Theoretical Physics, Physics Dept.,
Univ. of Michigan, Ann Arbor, MI 48109\\ 
$^2$Physics Dept., University of Utah, Salt Lake City, UT\\
$^3$Dept.\ of Physics and Astronomy, Rutgers University, Piscataway, NJ 08854\\
$^{4}$Physics Dept., University of California, Santa Cruz, CA 95060
\end{center}

\begin{abstract}
\noindent
The first stars in the universe form inside $\sim 10^6 M_\odot$ dark
matter (DM) haloes whose initial density profiles are laid down by
gravitational collapse in hierarchical structure formation scenarios.
During the formation of the first stars in the universe, the baryonic
infall compresses the dark matter further.  The resultant dark matter
density is presented here, using an algorithm originally developed by
Young to calculate changes to the profile as the result of adiabatic
infall in a spherical halo model; the Young prescription
takes into account the non-circular motions of halo particles.
The density profiles
obtained in this way are found to be within a factor of two of those
obtained using the simple adiabatic contraction prescription of
Blumenthal et al.  Our results hold regardless of the nature of the
dark matter or its interactions and rely merely on gravity.  If the
dark matter consists of weakly interacting massive particles, which
are their own antiparticles, their densities are high enough that
their annihilation in the first protostars can indeed provide an
important heat source and prevent the collapse all the way to fusion.
In short, a ``Dark Star'' phase of stellar evolution, powered by DM
annihilation, may indeed describe the first stars in the universe.

\end{abstract}

\section{Introduction}
The first stars in the Universe mark the end of the cosmic dark ages,
reionize the Universe, and provide the enriched gas required for later
stellar generations.  They may also be important as precursors to
black holes that coalesce and power bright early quasars.  The first
stars are thought to form inside dark matter halos of mass $ 10^5
M_\odot$--$ 10^6 M_\odot$ at redshifts $z=10-50$.  These halos consist
of 85\% dark matter and 15\% baryons in the form of metal-free gas made of
hydrogen and helium.  Theoretical calculations indicate that the
baryonic matter cools and collapses via molecular hydrogen cooling
\cite{Peebles:1968nf,matsuda,Hollenbach} into a single small protostar
\cite{omukai} at the center of the halo (for reviews see
e.g. \cite{Ripamonti:2005ri,Barkana:2000fd,Bromm:2003vv}).

Previously, three of us\cite{sfg} (hereafter, Paper I) considered the
effect of dark matter (DM) particles on the first stars during their
formation.  We focused on Weakly Interacting Massive Particles
(WIMPs), which are the favorite dark matter candidate of many
physicists because they automatically provide approximately the right
amount of dark matter, i.e.\ 24\% of the current energy density of the
Universe.  WIMPs are their own antiparticles, and annihilate with
themselves in the early universe, leaving behind this relic density.
Probably the best example of a WIMP is the lightest supersymmetric
particle. In particular, the neutralino, the supersymmetric partner of
the W, Z, and Higgs bosons, has the required weak interaction cross
section and mass $\sim$ GeV - TeV to give the correct amount of dark
matter. For a review of SUSY dark matter see \cite{jkg}
\footnote{This same annihilation process is important wherever the
  WIMP density is sufficiently high.  Such regimes include the early
  Universe, in galactic halos today \cite{ellis,gs}, in the Sun
  \cite{sos} and Earth \cite{freese,ksw}, and in the first stars.}.
As our canonical values, we used the standard value $\langle \sigma v
\rangle = 3 \times 10^{-26}\,{\rm cm^3/s}$ for the annihilation cross
section and $m_\chi = 100\,{\rm GeV}$ for the WIMP particle mass, but
also considered a broader range of WIMP masses (1 GeV--10 TeV) and
cross-sections.\footnote{ The interaction strengths and masses of the
  neutralino depend on a large number of model parameters. In the
  minimal supergravity model, experimental and observational bounds
  restrict $m_\chi$ to 50 GeV--2 TeV, while $\langle \sigma v \rangle$
  lies within an order of magnitude of $3 \times 10^{-26}\, {\rm
    cm^3/sec}$ (except at the low end of the mass range where it could
  be several orders of magnitude smaller).  Nonthermal particles can
  have annihilation cross-sections that are many orders of magnitude
  larger (e.g.~\cite{moroi}) and would have even more drastic effects.
  With present state of the field there are many types of DM
  candidates which could apply\cite{fargion}}

We found that a crucial transition takes place when the hydrogen density of
the collapsing protostar exceeds a critical value ($10^{13}\, {\rm
cm}^{-3}$ for a 100 GeV WIMP mass): at this point WIMP annihilation
heating dominates over all cooling mechanisms and prevents the further
collapse of the star. We suggested that the very first stellar objects
might be ``Dark Stars,'' a new phase of stellar evolution in which the
DM -- while only supplying $\sim 1$\% of the mass density -- provides the
power source for the star through DM annihilation.  We are as yet
uncertain of the lifetime of these theoretical objects.  It is
possible that they last as long as 600 Myr, which is the timescale for
all the DM to annihilate away inside of the Dark Star (neglecting
refilling the region using DM from the halo outside of the Dark Star).
This lifetime is as yet uncertain.

The existence of Dark Stars depends on the fact that the DM densities
in these protostars is sufficiently high for the DM annihilation
heating to be significant. Previously, we obtained estimates for the
DM densities as follows: we started with a variety of initial density
profiles laid down by gravitational collapse during hierarchical
structure formation (e.g., Navarro, Frenk, and White profiles).  These
profiles were then compressed further due to the baryonic quasi-static 
                            contraction as
the protostar cooled and collapsed.  As the baryons come to dominate
the potential well in the core, they pull the DM particles inward.
Previously we used the simple adiabatic contraction method
of Blumenthal {\it etal} \cite{Blumenthal:1985qy,barneswhite,rydengunn}
(hereafter Blumenthal method) to estimate the resultant DM density profile.
This method is extremely simplistic, in that it assumes all the halo
particles move on circular orbits; equivalently only their angular
momentum is conserved as the halo is compressed.  We found that this
method gave a DM density at the outer edge of the baryonic core
\begin{equation}
\label{eq:blum}
\rho_\chi \simeq 5 {\rm GeV/cm}^{-3} (n/{\rm cm}^{3})^{0.81} 
\end{equation}
and scales as $\rho_\chi \propto r^{-1.9}$ outside the core.  However,
a number of authors have compared the Blumenthal method to
$N$-body simulations of compressed haloes in other contexts (galaxies),
and warned that the predicted density profiles in those cases were
more concentrated than found in their $N$-body simulations, especially
in the crucial inner part \cite{barnes,sell99,gnedin04}.  Consequently
there was some concern whether or not the DM densities used in our
first paper on Dark Stars were badly overestimated, so that it
remained unclear whether or not DM heating in the first protostellar
objects is important.

It is the purpose of this paper to show that the DM density in the
central core of the protostars does indeed grow high enough for the DM
heating to overpower any cooling mechanism, thereby preventing the
further collapse of the star and giving rise to a dark star powered by
DM annihilation.  We follow the approach of Young (1980) who presented
a general treatment of adiabatic compression of a spherical system.
His formulation was originally to model the growth of a black hole in
a spherical star cluster, but the method can be applied for any
adiabatic change to a spherical potential.  It has been used to study
halo compression by Wilson \cite{wilson} and independently by Sellwood
and McGaugh\cite{sellmcgaugh} in galaxies.  The latter authors found
that the simplistic Blumenthal method overestimated the central
densities in galaxies by factors of two or three, but {\bf not} by
several orders of magnitude, even when radial pressure and disk
geometry are taken into account.  Here we apply Young's method, as
adapted by Sellwood and McGaugh, to protostellar Pop III objects in
order to obtain the DM densities inside their cores.

It is important to point out that Young's adiabatic prescription is
not at all sensitive to departures from sphericity - Sellwood and
McGaugh \cite{sellmcgaugh} showed that, 
 for what concerns adiabatic contraction, even a flat disk could be well
approximated as a sphere.

We briefly mention here other previous work on DM annihilation in
stars.  Previous work on DM annihilation powering stars has also been
done in the context of high DM densities near the supermassive black
holes in galactic centers, e.g. WIMP burners \cite{moskalenko} and
more generally \cite{edsjo,bertone2007}.  In another paper, we
considered the effects of DM annihilation on early zero-metallicity
(Population III) stars, once they do have fusion inside their
cores\cite{fsa}.  A very similar work was simultaneously submitted by
\cite{iocco}.  These stars live inside a reservoir of WIMPs; as the
WIMPs move through the stars, some of the WIMPs are captured by the
stars.  The captured DM sinks to the center of the stars, where it
adds to the DM annihilation.  We found that DM annihilation may
dominate over fusion, and the DM power source may exceed the Eddington
luminosity and prevent the first stars from growing beyond a limited
mass.

The DM profiles obtained in this paper are appropriate to {\it any}
type of collisionless DM, not just WIMPs.  This work relies upon the
gravitational response of the halo to the baryons alone, and requires
only that the DM particles are both collisionless and interact with
the baryons only through gravitational forces.  We also need to make
an assumption about the velocity distribution of the DM particles in
the very centers of these small halos, which we take to be isotropic.

The protostellar clouds (unlike galaxies as a whole) are baryon rich
environments.  For hydrogen density $n> 10^4$cm$^{-3}$, the baryon
density exceeds the DM density in the collapsing protostar.  We note
that the angular momentum of the collapsing baryons is transferred to
baryons via gas dynamics rather than being transferred to dark matter
gravitationally.  In Section 2, we will review the various methods for
halo compression.  In Section 3, we present our results in the first
protostellar objects.  Then we conclude in Section 4.

\section{Methods to Obtain the Dark Matter Density}
In this paper we assume adiabaticity of the protostellar collapse.  In
other words, since the change of the gravitational potential is not
too substantial over a single orbital time scale during the phase we
are studying, we may take the adiabatic invariants to be well
conserved.

\subsection{Initial Dark Matter Density Profiles: NFW and Core}

With this assumption of adiabaticity, we follow the response of the DM
to the collapse of the protostellar object.  As our initial conditions
for the DM before contraction, we take an overdense region of
$10^5-10^6 M_\odot$ with a Navarro, Frenk and White (NFW) profile
\cite{NFW} for both DM and gas, where the gas contribution is 15\% of
that of the DM.  The density profile of an NFW halo is
\begin{equation}
\label{eq:NFW}
\rho(r) = {\rho_0 \over x(1+x)^2}
\end{equation}
where $x=r/r_s$, $r_s$ is the break radius of the profile and $\rho_0
\equiv 4 \rho(r_s)$ sets the density scale.  The critical density of
the universe varies with redshift $z$ as
\begin{equation}
\label{eq:rhocrit}
\rho_c(z) = {3 H^2(z) \over 8 \pi G} .
\end{equation}
The Hubble parameter in a flat universe varies as
\begin{equation}
\label{eq:hubble}
H^2(z) = H_0^2 [\Omega_{m} (1+z)^3 + \Omega_{\Lambda}]
\end{equation}
where the current matter
density is $\Omega_{m} \sim 0.25$, and the $\Omega_{\Lambda}$ term (dark energy) on
the right hand side can be neglected at the redshifts we are
considering.  We find $H^2(z=19) \sim 2000 H_0^2$, where today's Hubble
constant $H_0 = 100 h {\rm km/sec/Mpc}$ with $h^2 \sim 0.5$.  For
an NFW halo of mass $M_{200} = 7 \times 10^5 M_\odot$, and taking
\begin{equation}
{3 M_{200} \over 4 \pi r_{200}^3} = 200 \rho_c(z) ,
\end{equation}
gives $r_{200} \sim 123$pc.  For a value of the concentration parameter 
$c \equiv r_{200}/ r_s = 2$, we have $r_s = 61.5$ pc.

For comparison, we also use the extreme case of a profile which has a
constant density DM core in the inner region before contraction.

Given these initial profiles for the DM, we can follow
their response to the changing baryonic gravitational potential as the
protostellar gas condenses.
The gas density profiles during the protostellar collapse we use are
taken from simulations of \cite{ABN,Gao06} (see Fig. 2 of
\cite{ABN}).  Roughly, the profiles of the hydrogen number density
have the functional form $n(r) = {n_{core} \over 1 + (r/r_B)^{2.3}}$.
The profile is flat from the origin out to (almost) $r_B$, where it
turns over to $n \sim r^{-2.3}$.  The core densities $n_{core}$ and
turnover radii $r_B$ are taken from \cite{ABN,Gao06}. We note that
the hydrogen number density should be multiplied by $\sim 1.35$ to
obtain the total gas density, in order to 
account for the additional presence of helium in the collapsing
protostellar cloud.

Below we will describe the results of adiabatic contraction for the DM
that we previously obtained using the Blumenthal method, and then turn
to the more accurate Young method to find better DM density profiles
in the collapsing protostellar objects.

\subsection{Previous estimates of DM density profile using Blumenthal method}
We previously used the Blumenthal method for adiabatic contraction
\cite{Blumenthal:1985qy,barneswhite,rydengunn} to obtain estimates of
the DM profile as the first protostars condensed.  As mentioned
previously, this method assumes that DM particles conserve 
angular momentum as the halo is compressed (other adiabatic
invariants set to zero); i.e., all particles are on
circular orbits.  One can assume that a particle at radius $r_i$ is
pulled in to a radius $r_f$.  Then one finds that $M_f(r_f) r_f =
M_i(r_i) r_i$, where $M(r)$ is the total mass inside radius $r$
and final quantities use the perturbed potential
(due to baryonic infall) for their evaluation.  Additionally,
the assumption is made that dissipationless particle orbits do not
cross (since particles are on circular orbits). After contraction, we
found the DM density at the outer edge of the baryonic core is roughly
given by Eqn.(\ref{eq:blum}) and scales as $\rho_\chi \propto
r^{-1.9}$ outside the core.  Our results are shown in Figure 1.

\begin{figure}[t]
\centerline{\includegraphics[width=0.5\textwidth]{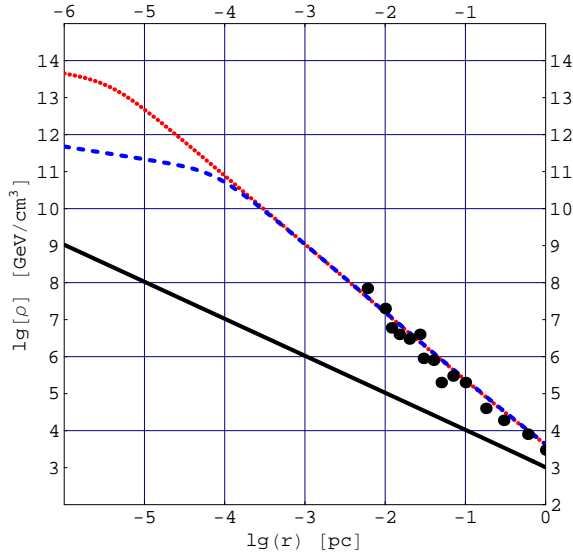}}
\caption{ Adiabatically contracted DM profiles for the first
collapsing protostars using the Blumenthal
method for an initial NFW profile (solid black line).  Resulting
curves correspond to hydrogen densities of $n =10^{13} {\rm cm}^{-3}$
(blue dashed lines) and $n = 10^{16}$cm$^{-3}$ (red dotted lines).
For comparison, the black dots show the DM densities obtained in the
numerical simulations of \cite{ABN} (see their Figure 2A)
corresponding to $n =10^{13}$cm$^{-3}$.  Excellent agreement is seen;
the outer profile for all cases scales as $\rho \propto r^{-1.9}$.
\vspace{-\baselineskip}
}
\end{figure}

One can compare our adiabatically contracted NFW profiles obtained via
the Blumenthal method with the DM profiles found numerically
in~\cite{ABN} for the first collapsing protostellar clouds 
(see their Fig.2).  While these authors obtained
remarkably good density profiles for the collapsing protostellar gas
in to very small radii, they found DM density profiles only up to
radii that are several orders of magnitude larger in extent.  In their
paper they presented two DM profiles (their earliest and latest
profiles), for $n \sim 10^3 {\rm cm}^{-3}$ and $n \sim 10^{13} {\rm
cm}^{-3}$ as far inward as $0.1$pc and $5 \times 10^{-3}$pc
respectively.  The slope of these two curves is the same as ours,
$\rho_\chi \propto r^{-1.9}$.  If one uses this slope and extrapolates
inward to smaller radii and to higher densities, then one obtains the
same DM densities as with our adiabatic contraction approach, as seen
in Figure 1.  Although we are encouraged by this agreement, one might
still worry about the extrapolation of the DM in toward radii smaller
than the ones treated by \cite{ABN}.

Hence it is the purpose of this paper to improve upon the Blumenthal
approximations used in Paper I and to obtain better estimates of the
DM density profiles inside the collapsing baryonic core in the first
protostars.   We note that $N$-body simulations (without assuming
adiabaticity) to further confirm the DM profiles obtained in this work
after baryonic compression are in progress \cite{zemp}.

\subsection{New Estimate of DM Density using Young's method} 
As discussed in Section 3.6 of Binney and Tremaine \cite{bt},
adiabatic changes conserve all three actions of an orbit.  In this
paper we do indeed work with the assumption that the growth of the
protostellar object is adiabatic.  In general, if no further
assumptions are made, one needs to consider all three actions, and
the problem becomes intractable analytically, though it can still be
followed with $N$-body simulations.  
In a spherical system, however, the problem reduces to two
conserved actions and becomes tractable: these two actions are the
angular momentum and radial action, while the third action is
identically zero because the plane of each orbit is an invariant.
Thus adiabatic compression of a spherical halo needs to take account
of the conservation of radial action, as well as of angular momentum.
Whereas Blumenthal's method makes the further restriction of only
circular orbits so that only angular momentum is conserved, Young
\cite{young} described an algorithm to take into account conservation
of both of these two adiabatic invariants.  The central idea of his
method is to require the distribution function to be invariant during
adiabatic changes: $f_i(J_r,J_\phi) = f_n(J_r,J_\phi)$, where
$J_r(E,L)$ is the radial action and $J_\phi \equiv L$ is the azimuthal
action, or total angular momentum per unit mass. Here subscripts $i$
and $n$ refer to initial and new halo profiles, computed in the
changing potential well due to baryonic infall.

Sellwood and McGaugh\cite{sellmcgaugh} implemented Young's method for halo
compression in galaxies (see also Wilson \cite{wilson}).  In their
paper, they describe an iterative procedure for obtaining a DM density
profile in response to a baryonic mass inside the halo that condenses.
They focused on a growing disk inside a galaxy, and the consequent DM
particle compression in the central region of the galaxy.  
They showed that Young's prescription was successful (matched numerical
results) even for a flat disk; indeed the method is not restricted
in practice to spherical systems.  In the present
paper the code of Sellwood and McGaugh is applied to the problem of
the first protostars, with this same iterative procedure following the
compression of the DM halo as the protostellar core collapses.

The method works by relating the distribution function of the
compressed halo to that of the original halo through the two conserved
actions, assuming the compression is adiabatic.  It is therefore
necessary to compute (1) the radial action from the energy and angular
momentum in the compressed potential, and then (2) the energy in the
original halo for the same radial action and angular momentum.  A
numerically efficient procedure is to interpolate from tables of these
quantities, and the radial action table for step (1) has to be
recomputed at each iteration.  Since the first stars are much smaller,
relative to the NFW halo break radius, than the galaxy disks used in
the previous application, we increased the sizes of the tables to
$200^2$ values in order to achieve the desired precision.  The final
density profile generally converged satisfactorily in a few, typically
3 or 4, iterations.

\section{Results}

We find the adiabatic contraction of DM densities due to the
collapsing gas densities in the inner regions of the first collapsing
protostellar clouds.  Our results can be seen in Figures 2 and 3 as
well as Table 1.  

\subsection{Initial NFW Profile}

We can start from an initial NFW halo profile for both the DM and the
baryons, with the baryons comprising 15\% of the total mass.  Starting
from initial NFW haloes (dashed line in the figure), Fig. 2 plots
contracted density profiles using (a) the standard Blumenthal method
for adiabatic contraction (dotted lines) and (b) the Young method
(solid lines), for a variety of hydrogen densities, $n = 10^4, 10^8,
10^{13}$, and $10^{16} {\rm cm}^{-3}$ for a total DM halo mass of $7
\times 10^5 M_\odot$ and a concentration parameter $c=2$ at $z=19$.
The upper panel illustrates the dark matter density profile and the
lower panel illustrates the enclosed DM mass $M(r)$ as a function of
radius.  Both profiles are shown in to a radius of $10^{-4}$pc.

\begin{table*}
 \begin{tabular}{||l|l|l|c||}
 \hline\hline
 Radius(pc) & DM Density w/ Young method & DM Density w/ Blumenthal method \\
 \hline
     2.42 $\times 10^{-4}$
   & $1.17 \times 10^{10}$ GeV/cm$^{3}$
   & $2.14 \times 10^{10}$ GeV/cm$^{3}$ \\
 \cline{1-3} 
     $9.30 \times 10^{-4}$ 
   & $1.07 \times 10^9$ GeV/cm$^{3}$
   & $1.95 \times 10^9$ GeV/cm$^{3}$ \\
 \cline{1-3}
     $1.11 \times 10^{-2}$ 
   & $1.13 \times 10^7$ GeV/cm$^{3}$
   & $2.07 \times 10^7$ GeV/cm$^{3}$ \\
 \hline
     $0.108$ 
   & $1.75 \times 10^5$ GeV/cm$^{3}$
   & $3.16 \times 10^5$ GeV/cm$^{3}$ \\
 \hline
     $1.08$ 
   & $3.02 \times 10^3$ GeV/cm$^{3}$
   & $5.13 \times 10^3$ GeV/cm$^{3}$ \\
 \hline
 \hline\hline
 \end{tabular} 
 \caption{
Contracted DM densities versus radius using Young and Blumenthal methods
for hydrogen density $n=10^{13}$cm$^{-3}$.  Note that the two methods
agree to within a factor of two.
 }
 \vspace{-5pt}
 \label{tab:ExpParam}
\end{table*}

\begin{figure}[t]
\centerline{\includegraphics[width=0.9\textwidth]{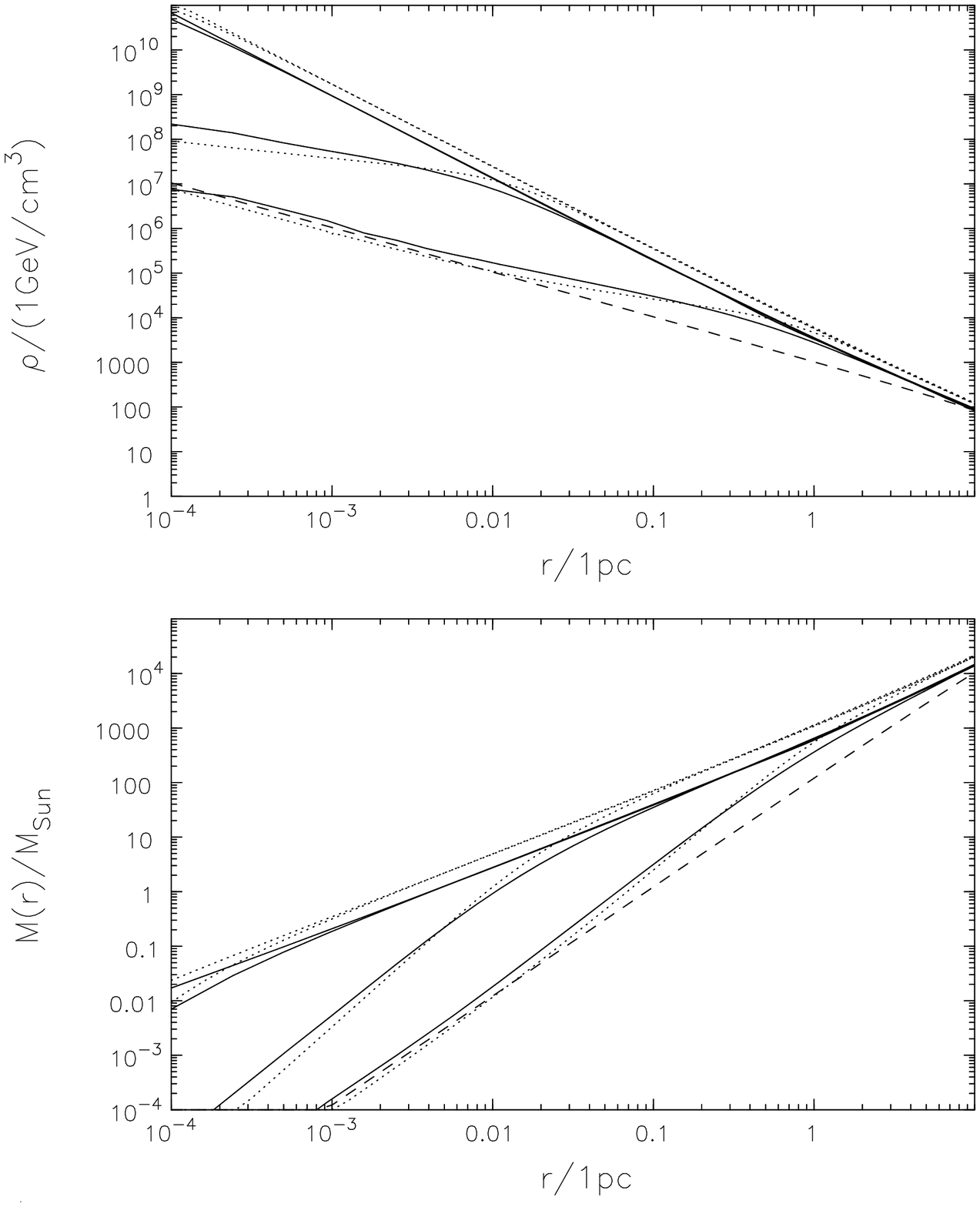}}
\caption{ Adiabatically contracted DM profiles in the first
protostars for an initial NFW
profile (dashed line) using (a) the Blumenthal method (dotted lines)
and (b) Young's method (solid lines), for $M_{\rm vir}=5 \times 10^7
M_\odot$, $c=2$, and $z=19$.  The upper panel shows the
resultant DM density profiles, and the lower panel shows the enclosed
DM mass as a function of radius.  The four sets of curves in each panel
correspond to a baryonic core density of $10^4, 10^8, 10^{13},$ and
$10^{16}{\rm cm}^{-3}$.  Our main result is that the two different
approaches to obtaining the DM densities find values that differ by
less than a factor of two.
\vspace{-\baselineskip}
}
\end{figure}

The main result of our paper is that the densities computed in both
 ways differ by no more than a factor of 2. In some cases the
 Blumenthal method yields a higher density, while in other cases the
 Young method yields a higher density; yet the difference never
 exceeds a factor of two.  One can see that the DM density is not
 enhanced at small radii for low gas density $n=10^4$cm$^{-3}$; this
 is because the amount of gas mass at these radii is simply too small.
 However, once the protostellar cloud gets to $n=10^8$cm$^{-3}$ or
 higher, the DM density is substantially enhanced due to adiabatic
 contraction using either method.  One might also ask why the
 Blumenthal formula yields a lower predicted density than Young's
 formula at small radii, as the plot shows for the two lower density
 stars.  The answer is that a halo particle on a non-circular orbit
 spends part of its time at large radii where the interior compressing
 mass is larger than when it is close to the center.  Thus the mean
 density can be increased even when little compressing mass is
 interior to that radius - this can happen only where the star has a
 diffuse core.

We note that the curves for $n=10^{16}$cm$^{-3}$ do not look
substantially different from those of $n=10^{13}$cm$^{-3}$. This is
because at the radius of $10^{-4}$pc, the two curves are just
beginning to diverge. At smaller radii, the higher gas density would
lead to substantially higher DM density; i.e., we expect that the DM
density for the $10^{16}$cm$^{-3}$ curve would be much higher than
that for the $10^{13}$cm$^{-3}$ curve at a radius of $10^{-6}$pc, but
we do not have the resolution to show this explicitly.  In short,
starting from an NFW profile, we have confirmed that the DM density in
the inner regions of the protostellar cloud is substantially enhanced
due to contraction during gas collapse.

\subsection{Initial Cored Profile}

Figure 3 plots the contracted density profiles obtained with the same
two techniques, but starting from a DM core as the initial
condition. To be extremely conservative, we take as our initial core
DM density $\rho_\chi \sim 3 \times 10^4$
(with a core radius of $0.2$pc).
This value is motivated by the DM density found by
\cite{ABN} (see Fig 2A) corresponding to a hydrogen density of
$10^3$cm$^{-3}$; their numerical results were found only in to a
radius $\sim 0.1$pc, and we simply take the value they found near that
radius to be constant inward all the way to the origin.  Clearly, this
gives an underestimate of the initial DM density in the central region
of the DM halo.  Even in this case, the DM density is substantially
enhanced by the subsequent gas collapse.  This initial profile is not
meant to be realistic and is presented merely as an extremely
conservative illustration of the two techniques for obtaining the DM
density profile.  In Figure 3, the three curves correspond to hydrogen
densities $n= 10^4, 10^8,$ and $10^{13} {\rm cm}^{-3}$.  The upper
panel illustrates the resultant DM profiles and the lower panel
illustrates the enclosed DM mass as a function of radius.  Again, the
DM density is substantially enhanced due to contraction during gas
collapse.  We note that, on the smallest scales and highest densities
(the $10^{13}$ cored case), the plots look irregular (particularly on
logarithmic scales) due to
a sparse numerical grid, which has fewer points in the inner region
for this cored case.

\begin{figure}[t]
\centerline{\includegraphics[width=0.9\textwidth]{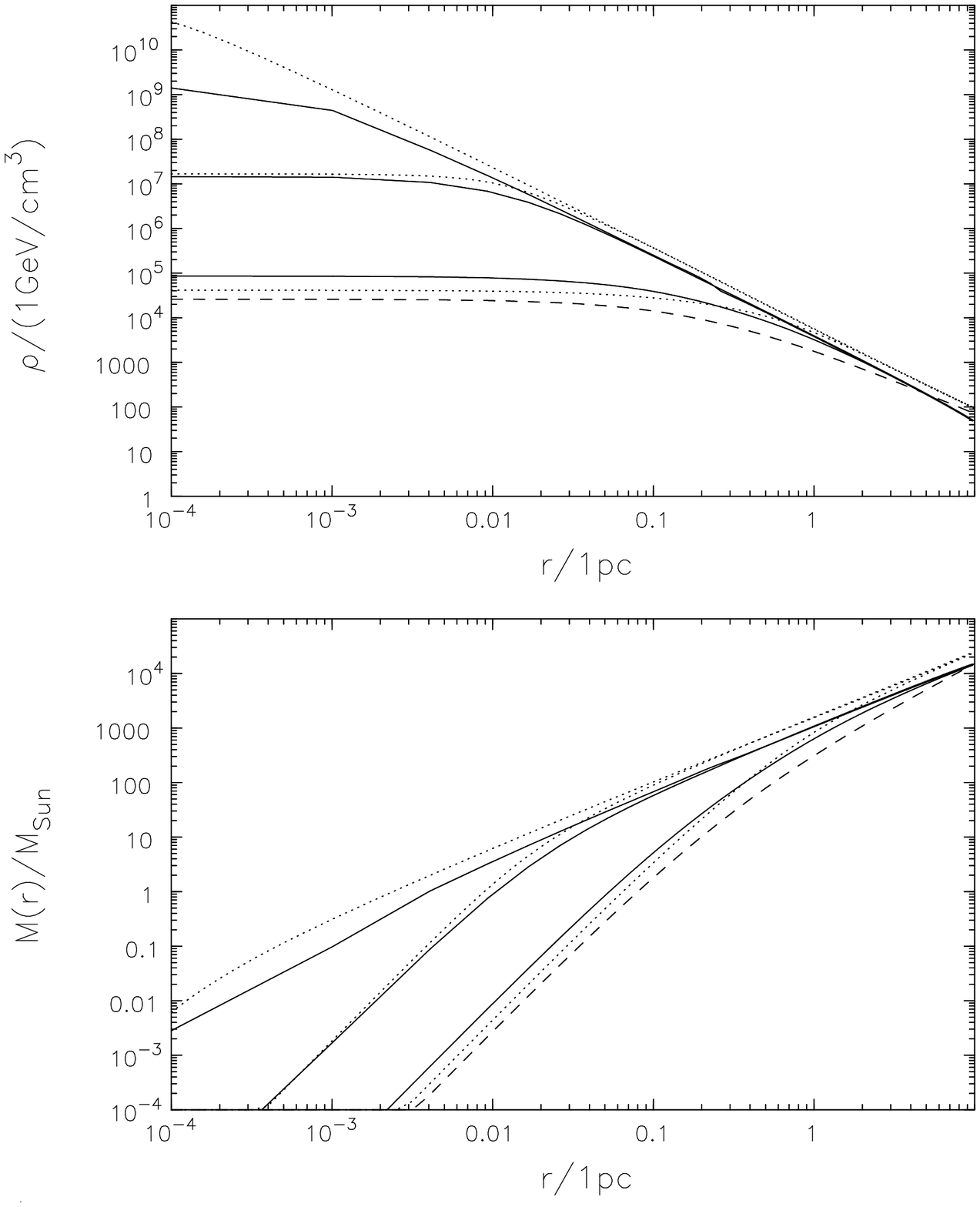}}
\caption{ Adiabatically contracted DM profiles and enclosed DM mass
for an initial cored profile (dashed line) using (a) the Blumenthal
method (dotted lines) and (b) Young's method (solid lines), with
parameters described in the previous figure.  The three sets of curves
in each panel correspond to a baryonic core density of $10^4, 10^8,$
and $10^{13}{\rm cm}^{-3}$.  This initial profile is not considered to
be realistic in dark matter haloes and is presented merely as an
illustration of the two techniques.  [The kink in the highest density
case at small radii using the Young method is not to be taken
seriously, as it is an artifact due to numerical uncertainties
significant only for this highest density cored case at small radii.]
\vspace{-\baselineskip}
}
\end{figure}

Our general conclusion is that DM densities are indeed substantially
enhanced in the presence of the collapsing baryonic protostellar
cloud.  Due to the resultant large DM densities, the DM heating
becomes substantial and the Dark Star proposed in Paper I does indeed
come into existence at this point.

\section{Discussion}
As the first stars form inside DM haloes, the protostellar baryons
gravitationally pull the DM farther inward.  Our results hold
regardless of the nature of the dark matter: simply due to gravity,
the density of any type of dark matter in the inner regions of the
collapsing Population III protostars will be enhanced by the baryonic
infall.  In this paper we computed the changes to the DM density
profile in the protostellar cores due to adiabatic increase of gas
densities in these cores.  We used Young's method appropriate for
spherical halos, which takes into account conservation of both angular
momentum and the radial action.  With this method we found DM density
profiles after contraction with central densities no more than a
factor of two different than those obtained previously \cite{sfg} via
a simple Blumenthal model for adiabatic contraction (which takes into
account only circular orbits).  The central DM densities in the
protostar are substantially enhanced during the gas collapse.  Note
that the compressed halo densities we obtain when we assume either
that halo particles have entirely circular orbits or an isotropic
velocity distribution do not differ by much; we do not therefore
expect our conclusions to be at all sensitive to any assumption for
the distribution of halo particle velocities.  $N$-body simulations
(without assuming adiabaticity) to further confirm the DM profiles
obtained in this work after baryonic compression are also in progress
\cite{zemp}.  We also reiterate that Young's adiabatic prescription is
not at all sensitive to departures from sphericity, as shown for a flat disk by  Sellwood and
McGaugh \cite{sellmcgaugh}.

If the DM consists of dark matter particles which annihilate among
themselves (the prototypical example is neutralino supersymmetric
particles), then these high DM densities lead to substantial DM
annihilation to the point where DM heating dominates in the stellar
core and prevents further collapse.  Indeed a Dark Star is created,
powered by DM annihilation rather than by fusion.

We acknowledge support from: the DOE and MCTP via the Univ.\ of
Michigan (K.F.); NSF grant AST-0507117 and GAANN (D.S.); NSF grant
PHY-0456825 (P.G.); NSF grant AST-0507323 (JS).  K.F. acknowledges the
hospitality of the Physics Dept. at the Univ. of Utah.

\end{document}